\documentclass[aps,prb,showpacs,twocolumn,superscriptaddress,floatfix]{revtex4-1}
\pdfoutput=1
\newcommand{\figurescale}{1}

\usepackage{bm}
\usepackage{verbatim}
\usepackage{amsmath}
\usepackage{amssymb}
\usepackage[utf8]{inputenc}
\usepackage[pdftex]{graphicx}
\usepackage[separate-uncertainty=true]{siunitx}
\DeclareSIUnit{\rpm}{rpm}
\usepackage[colorlinks=true,urlcolor=blue,linkcolor=blue,citecolor=blue]{hyperref}
\usepackage{xcolor}
\usepackage[T1]{fontenc}
\usepackage{placeins}
\usepackage{nicefrac}
\usepackage{braket}
\usepackage{pdfcomment}
\usepackage{xcolor}
\usepackage[euler]{textgreek}
\usepackage[normalem]{ulem}

\usepackage{pbox}
\usepackage{makecell}
\usepackage{hhline}
\usepackage{multirow}
\usepackage{tabularx}

\definecolor{greentwo}{RGB}{0,180,0}

\newcommand{\angstrom}{\text{\normalfont\AA}}

\begin{document}

\title{Electrical control of orbital and vibrational \textit{inter}layer coupling in bi- and trilayer 2H-MoS$_2$}
%
%
\author{J. Klein}
\affiliation{Walter Schottky Institut and Physik Department, Technische Universit\"at M\"unchen, Am Coulombwall 4, 85748 Garching, Germany}
\affiliation{Department of Materials Science and Engineering, Massachusetts Institute of Technology, Cambridge, Massachusetts 02139, USA}
\author{J. Wierzbowski}
\affiliation{Walter Schottky Institut and Physik Department, Technische Universit\"at M\"unchen, Am Coulombwall 4, 85748 Garching, Germany}
\author{P. Soubelet}
\affiliation{Walter Schottky Institut and Physik Department, Technische Universit\"at M\"unchen, Am Coulombwall 4, 85748 Garching, Germany}
\author{T. Brumme}
\affiliation{Wilhelm-Ostwald-Institute for Physical and Theoretical Chemistry, Leipzig University, Leipzig, Germany}
\affiliation{Theoretical Chemistry, TU Dresden, Dresden, 01062 Germany}
\author{L. Maschio}
\affiliation{Dipartimento di Chimica and Centre of Excellence NIS (Nanostructured Interfaces and Surfaces), Universit\`a di Torino, via P. Giuria 5, I-10125 Turin, Italy}
\author{A. Kuc}
\affiliation{Abteilung Ressourcen\"okologie, Helmholtz-Zentrum Dresden-Rossendorf, Forschungsstelle Leipzig, Leipzig, Germany}
\affiliation{Department of Physics \& Earth Science, Jacobs University Bremen, Bremen, Germany}
\author{K. M\"uller}
\affiliation{Walter Schottky Institut and Department of Electrical and Computer Engineering, Technische Universit\"at M\"unchen, Am Coulombwall 4, 85748 Garching, Germany}
\author{A. V. Stier}
\affiliation{Walter Schottky Institut and Physik Department, Technische Universit\"at M\"unchen, Am Coulombwall 4, 85748 Garching, Germany}
\author{J. J. Finley}
\affiliation{Walter Schottky Institut and Physik Department, Technische Universit\"at M\"unchen, Am Coulombwall 4, 85748 Garching, Germany}
%
%
\date{\today}
%
%

\begin{abstract}
\textbf{Manipulating electronic \textit{inter}layer coupling in layered van der Waals (vdW) materials is essential for designing opto-electronic devices. Here, we control vibrational and electronic \textit{inter}layer coupling in bi- and trilayer 2H-MoS$_2$ using large external electric fields in a micro-capacitor device. The electric field lifts Raman selection rules and activates phonon modes in excellent agreement with \textit{ab-initio} calculations. Through polarization resolved photoluminescence spectroscopy in the same device, we observe a strongly tunable valley dichroism with maximum circular polarization degree of $\sim 60\%$ in bilayer and $\sim 35\%$ in trilayer MoS$_2$ that are fully consistent with a rate equation model which includes input from electronic band structure calculations. We identify the highly delocalized electron wave function between the layers close to the high symmetry $Q$ points as the origin of the tunable circular dichroism. Our results demonstrate the possibility of electric field tunable \textit{inter}layer coupling for controlling emergent spin-valley physics and hybridization driven effects in vdW materials and their heterostructures.}
\end{abstract}

%
%
\maketitle
%
%

\section{Introduction}

Breaking of crystal symmetries has profound effects on the electrical~\cite{bilayer-gap,valley-current1,valley-current2} and optical properties~\cite{Wu.2013,spin-layer-locking,valley-Hall-MoS2,Klein.2016b} of vdW bonded materials and their heterostructures. A prime example is bilayer graphene, where an applied electric field breaks crystal inversion-symmetry that opens up a band gap~\cite{bilayer-gap} and induces, e.g., topological valley currents.~\cite{valley-current1,valley-current2} Monolayer transition-metal dichalcogenides (TMDCs), like MoS$_2$, exhibit valley dichroism, due to their inherently broken inversion symmetry, where $K$ valleys are related to one another via time-reversal symmetry, as governed by the Kramer's degeneracy.~\cite{Heinz-monolayer,Splendiani-monolayer,Heinz2012,Cui2012,Cao.2012,Sallen2012} The 2H stacking results in alternating point groups with odd layer numbers manifesting in broken spatial inversion symmetry that is restored in either even layer crystals or evenly layered crystals. Several intriguing experimental observations, including electrical tuning of valley-magnetic moment~\cite{Wu.2013}, spin-layer locking,~\cite{spin-layer-locking} and finite valley Hall effect,~\cite{valley-Hall-MoS2} have been reported in bilayer TMDCs and ascribed to the control of Berry curvature, due to inversion symmetry breaking by an external electric field.~\cite{Kormnyos.2018}

TMDCs exhibit a complex and intricate energy landscape in reciprocal space away from the $K$ points.~\cite{Steinhoff.2014,Steinhoff.2015,Steinhoff.2016} Other high symmetry points become relevant due to their momentum indirectness creating dark states that are experimentally not easily accessible, but contribute significantly to their optical and electronic properties.~\cite{Wu.2015,Qiu.2015,Selig.2016,Malic.2018,Lindlau.2018,Gerber.2019,Paradisanos.2020,Christiansen.2019,Mado.2020} In this context, the $Q$ point (also commonly denoted as $\Lambda$ or $\Sigma$ point) is of considerable importance due to the small energy difference between $K$ and $Q$ in the conduction band.~\cite{Steinhoff.2014} Time-resolved $\mu$-ARPES experiments have demonstrated that electrons can very effectively distribute between the $K$ and $Q$ points~\cite{Steinhoff.2014,Steinhoff.2016} as experimentally observed in monolayer MoS$_2$~\cite{Wallauer.2016} and other TMDCs.~\cite{Mado.2020} Furthermore, in multilayer TMDCs, the wave function overlap at the $Q$ point is more pronounced due to the dominant sulfur $p$-orbital admixture manifesting in efficient charge transfer, e.g., in heterobilayers.~\cite{Wang.2017,Jin.2018,Kunstmann.2018,Yuan.2020,Ramzan.2021} 

In such multilayered systems, \textit{inter}layer hopping of electrons or holes plays a crucial role for the formation of \textit{inter}layer excitonic complexes.~\cite{Miller.2017,Wang.2017,Jin.2018,Gerber.2019,Paradisanos.2020,Leisgang.2020,Lorchat.2021} The re-distribution of charge carriers is realized by an ultra-fast charge transfer between the layers.~\cite{Hong.2014,Jin.2018} Here, the application of an external electric field can control the layer hybridization~\cite{Kiemle.2020,Gao.2017,Shimazaki.2020,Leisgang.2020} and tune hopping rates between individual layers to control photo-physical properties. Therefore, realizing field controlled multilayer devices made from TMDCs may offer novel device designs to control, e.g. spin-valley polarization for prospective opto-electronic devices.

In this Letter, we electrically control the \textit{inter}layer hybridization in bi- and trilayer MoS$_2$. From our measurements, we observe electric field-activated Raman modes, where the relative mode intensity is a fingerprint for the degree of \textit{inter}layer (de)coupling. The observed activated phonon modes are in excellent agreement with $ab$-$initio$ DFT calculated phonon spectra that include external electric fields. The phonon mode activation correlates with a highly tunable degree of circular polarization $\eta$ of the $A$-exciton emission in inversion symmetric bilayer MoS$_2$, with values varying with the electric field from $\eta \sim 0.2$ up to $0.6$, limited only by the maximally explored gate-voltage. Strikingly, the trilayer sample, which is intrinsically inversion symmetry broken, also reveals a tunable $\eta \sim 0.2$ to $0.36$, unlike the monolayer MoS$_2$, where $\eta$ is field-independent. We interpret our observations in terms of the electric field dependent conduction and valence band edges at the $K$/$K^{\prime}$ and $Q$/$Q^{\prime}$ points. This changes the \textit{inter}layer hybridization, most importantly for our experiments the electron wave function distribution between the layers at the $Q$/$Q^{\prime}$ points results in an electric field dependent population of optically excited electrons and holes in the layers at the $K$/$K^{\prime}$ points. We support our observations with a rate-equation model that takes into account the $ab$-$initio$ calculated electronic band structure. Our results show that the $Q$ points significantly influence electron hopping in 2H-multilayered TMDC crystals, and that the concept of symmetry alone is insufficient to explain optical dichroism in few-layer MoS$_2$ but that \textit{inter}layer coupling and in particular, the hopping of electrons strongly influences the redistribution of optically excited carriers in the multilayer sample. Our results further demonstrate the potential to control emergent spin- and valleytronic devices based on two-dimensional (2D) atomically thin crystals with electric fields. 

\section{Results and Discussion}

\subsection{Microcapacitor device}

\begin{figure*}
	\scalebox{\figurescale}{\includegraphics[width=1\linewidth]{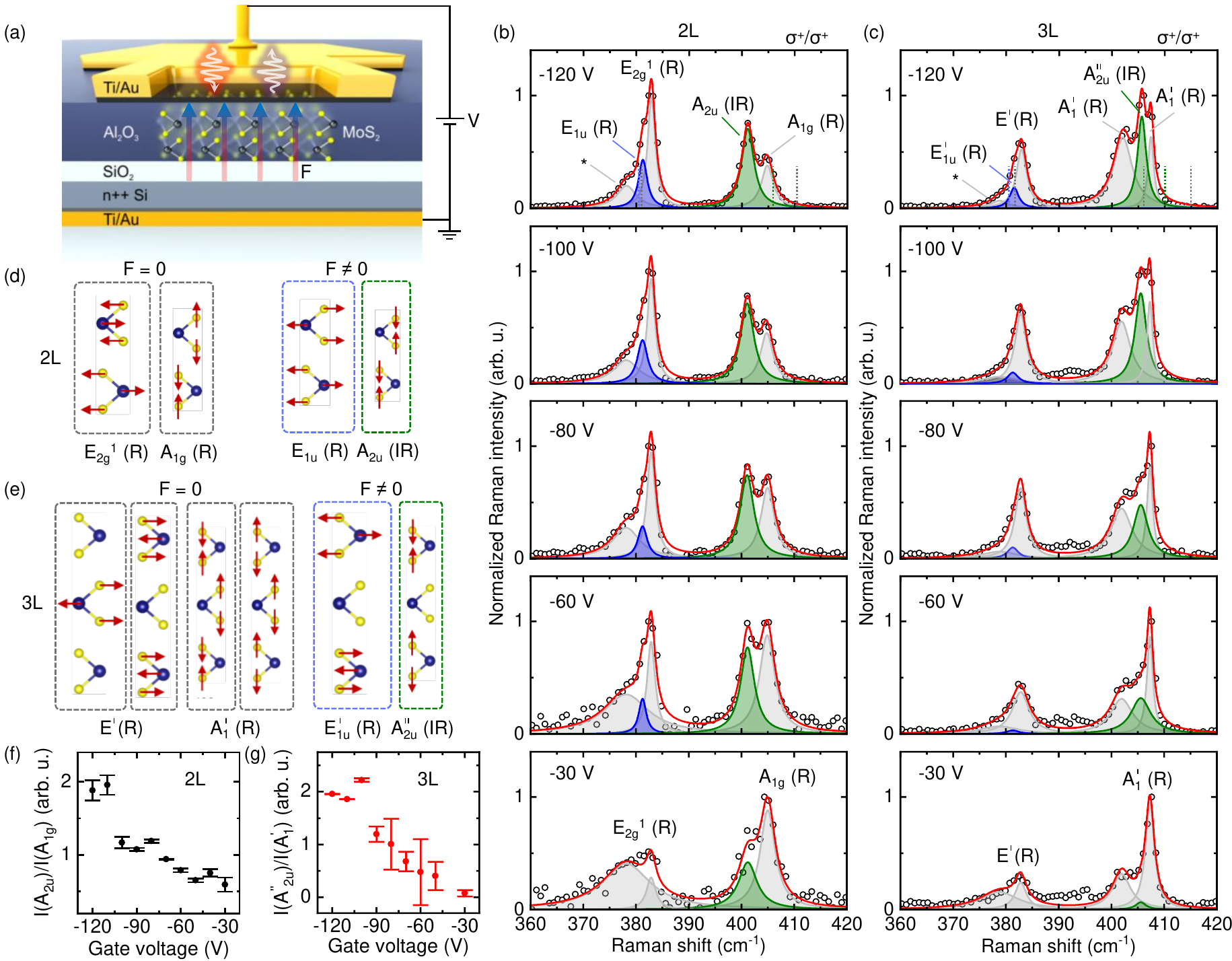}}
	\renewcommand{\figurename}{FIG.|}
	\caption{\label{fig1}
		(a) Schematic illustration of the micro-capacitor device. Few-layer MoS$_{2}$ is electrically isolated due to encapsulation between SiO$_2$ and Al$_2$O$_3$. The $\SI{5}{\nano\meter}$ thick Ti top-gate allows optical access while out-of-plane electric fields are applied.
		(b) Voltage dependent low-temperature ($\SI{10}{\kelvin}$) Raman spectra for bilayer MoS$_2$. Breaking of crystal symmetry from the external electric field activates Raman modes $E_{1u}$ (R) (blue) and $A_{2u}$ (IR) (green).
		(c) Voltage dependent Raman spectra for trilayer MoS$_2$. The electric field activates the Raman modes $E^{\prime}_{1u}$ (R) (blue) and $A^{\prime\prime}_{2u}$ (IR) (green). Dashed lines highlight corresponding calculated phonon frequencies.
		(d) Atomic displacement of Raman modes in bilayer and (e) trilayer MoS$_2$ with and without electric field. Dashed lines highlight corresponding calculated phonon frequencies.
		(f) Ratio of the Raman intensities of electric field activated phonon modes $A_{2u}$ and $A_{1g}$ indicate a measure of vibrational symmetry breaking in the bilayer.
		(g) Ratio of the Raman intensities of electric field activated IR Raman modes $A^{\prime\prime}_{2u}$ and $A^{\prime}_{1}$ in the trilayer sample.
		}
\end{figure*}

The investigated device consists of mono-, bi- and trilayer 2H-MoS$_2$ embedded in a micro-capacitor structure. The layers are electrically isolated from the contacts, as depicted schematically in Fig.~\ref{fig1}(a). A \SI{5}{\nano\meter} thick semi-transparent titanium top gate facilitates optical access to the crystals while applying a gate-voltage in the range of $\pm\SI{120}{\volt}$ results in electric fields on the order of $\SI{}{\mega\volt\per\centi\meter\square}$.~\cite{Klein.2016,Klein.2016b} Raman and PL data are acquired keeping the device at a lattice temperature of $\SI{10}{\kelvin}$ and optically exciting at an energy of $\SI{1.96}{\electronvolt}$.

\subsection{Activation of Raman modes by crystal symmetry breaking}

The lattice vibrational modes in 2D materials encode rich information of crystal symmetries that can be directly derived from group theory.~\cite{Verble.1970,MolinaSnchez.2011} The irreducible representation of a monolayer MoS$_2$ ($D_{3h}$ point group) is $\Gamma = 2A_{2}^{\prime\prime} \otimes A_{1}^{\prime} \otimes 2E^{\prime} \otimes E^{\prime\prime}$ with $2A_{2}^{\prime\prime} \otimes E^{\prime}$ representing the in-plane (LA/TA) and out-of-plane (ZA) acoustic phonon modes, $2A_{2}^{\prime\prime}$ the IR active mode, $A_1^{\prime}$ and $E^{\prime\prime}$ the Raman active modes, and $E^{\prime}$ representing an IR and Raman active mode.~\cite{Zhao.2013} Both, $E^{\prime}$ and $A^{\prime}_{1}$ Raman modes are active in our backscattering geometry. In order to study the effect of perturbing crystal symmetry using an out-of-plane electric field, we now tune the gate-voltage and study phonon modes.

\begin{table}
\begin{ruledtabular}
\begin{tabularx}{\linewidth} {cccc}

                    & Mode                & $\omega_{exp}$ ($\SI{}{\per\centi\meter}$) & $\omega_{theo}$ ($\SI{}{\per\centi\meter}$) \\  \hline 
\multirow{4}{*}{2L} & $E_{1u} (R)$        & $381.24\pm0.40$                            & $381.5$                                     \\
                    & $E_{2g}^{1} (R)$     & $382.81 \pm 0.05$                          & $381.7$                                     \\
                    & $A_{2u} (IR)$        & $401.17 \pm 0.08$                          & $407$                                       \\
                    & $A_{1g} (R)$         & $404.81 \pm 0.14$                          & $412$   \\  \hline  
\multirow{5}{*}{3L} & $E^{\prime}_{1u} (R)$        & $381.5\pm0.83$                             & $380.9$                                     \\
                    & $E^{\prime} (R)$     & $382.91\pm0.15$                            & $381.4$                                     \\
                    & $A^{\prime}_{1} (R)$ & $402.05 \pm 0.11$                          & $406$                                       \\
                    & $A^{\prime\prime}_{2u} (IR)$        & $405.67 \pm 0.05$                          & $410$                                       \\
                    & $A^{\prime}_{1} (R)$ & $407.39 \pm 0.05$                          & $415$    
                    \\
\end{tabularx}
\caption{Experimental and theoretical frequencies of phonon modes of bi- and trilayer MoS$_2$ with an external electric-field.}
\label{table:values_modes}
\end{ruledtabular}
\end{table}

Gate-voltage dependent data for bi- and trilayer MoS$_2$ are presented in Fig.~\ref{fig1}(b) and (c). For maximally broken inversion symmetry in the bilayer and also in the trilayer ($V = \SI{-120}{\volt}$ in our device) additional phonon modes (colored blue and green) appear in the spectrum (see Fig.~\ref{fig1}(b) and (c)). Their intensity monotonically decreases with decreasing gate-voltage.

To understand the origin of the field-activated phonon modes, we perform $ab$-$initio$ DFT calculations of phonon spectra for mono-, bi- and trilayer MoS$_2$ with and without electric field (see full phonon dispersion in the Supplemental Material). Our calculations indeed suggest additional electric-field-activated phonon modes, $E_{1u}$ (R) ($E^{\prime}_{1u}$ (R)) and $A_{2u}$ (IR) ($A^{\prime\prime}_{2u}$ (IR)) in the bilayer (trilayer), due to the lifting of Raman selection rules. Corresponding atomic displacements are shown in Fig.~\ref{fig1}(d) and (e). The experimental and theoretical values of phonon modes of bi- and trilayer are summarized in Tab.~\ref{table:values_modes} and highlighted in the Raman spectra in Fig.~\ref{fig1}(b) and (c). The calculated absolute phonon mode energies and their frequency differences are in excellent agreement with our experimental spectra. To quantify the extent to which symmetry is broken by the external electric field, we plot the ratio of phonon mode intensities between the $A_{2u}$ and $A^{\prime}_{1}$ modes for the bilayer MoS$_2$ and $A^{\prime\prime}_{2u}$ and $A^{\prime}_{1}$ modes for the trilayer MoS$_2$ (see Fig.~\ref{fig1}(f) and (g)). As expected, the highest ratio (degree of symmetry breaking) is observed for the maximum gate-voltage $V = \SI{-120}{\volt}$ (electric field) while the ratio vanishes for decreasing gate-voltages $V = \SI{-30}{\volt}$. The offset from $V = \SI{0}{\volt}$ originates from the asymmetric dielectric environment and built-in electric field of the device as discussed in detail in Refs.~\cite{Klein.2016,Klein.2016b}. Note that we focus our discussion on positive gate-voltages since the total Raman intensity strongly diminishes at negative voltages due to carrier doping effects in the asymmetric device structure,~\cite{Miller.2019} which originate from charge transfer for this field polarity from the Al$_2$O$_3$ interface into the MoS$_2$.~\cite{Klein.2016,Klein.2016b} We can conclude that the appearance of new Raman modes for an applied electric field is a direct fingerprint for the change in vibrational coupling in our tunable micro-capacitor device.

\subsection{Electrical control of \textit{inter}layer hybridization}

We now directly correlate the observed crystal symmetry breaking from our Raman measurements with circularly-polarized $\mu$-PL of 1L, 2L and 3L MoS$_2$. All measurements are performed with quasi-resonant excitation of the $A$-exciton using a CW HeNe laser emitting at $\SI{1.96}{\electronvolt}$ and the samples at a lattice temperature of $T = \SI{10}{\kelvin}$. Polarization resolved PL spectra are recorded with $\sigma^{+}$ (and $\sigma^{-}$) polarized excitation and the emitted PL is analyzed with $\sigma^+$ and $\sigma^{-}$ discrimination. Both sets of data are equivalent since the $K$ and $K^{\prime}$ valleys are equivalent. The resulting PL spectra are denoted $I^{\pm}(E)$, where the superscript denotes the helicity of the polarization discrimination in the detection channel and $E$ is the energy of the emitted photons.

\begin{figure}
	\scalebox{\figurescale}{\includegraphics[width=1\linewidth]{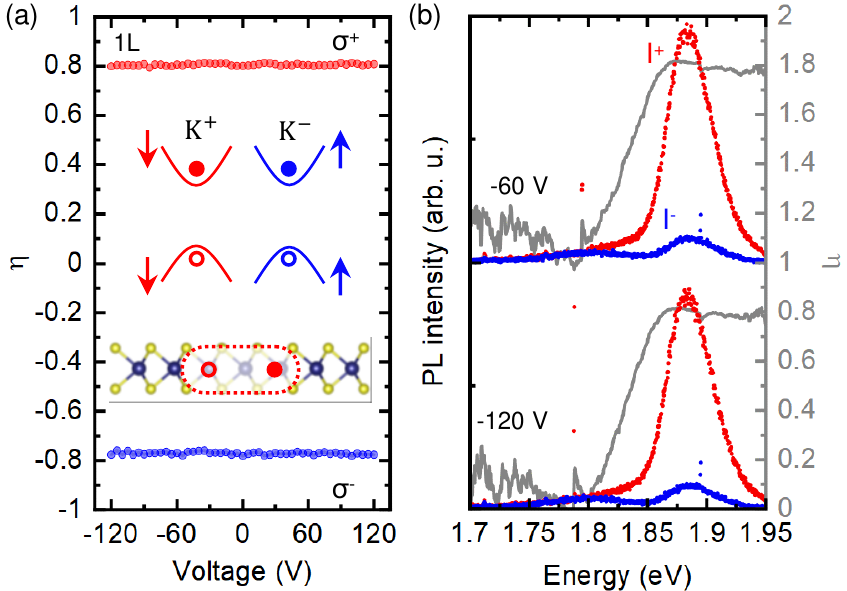}}
	\renewcommand{\figurename}{FIG.|}
	\caption{\label{fig2}
		(a) Gate-voltage independent degree of circular polarization $\eta$ for monolayer MoS$_2$ with $\sigma^{+}$ (red) and $\sigma^{-}$ (blue) excitation. Inset: Valley and spin \textit{inter}band optical selection rules at $K$ and $K^{\prime}$ and \textit{intra}layer exciton.
		(b) Low-temperature circularly polarized $\mu$-PL spectra for 1L taken at $\SI{-60}{\volt}$ and $\SI{-120}{\volt}$. Blue (red) spectra are obtained for $\sigma^{+}$ excitation and $\sigma^{+}$ ($\sigma^{-}$) detection. The degree of valley polarization is plotted in grey.
		}
\end{figure}

Typical gate-voltage dependent degree of valley polarization $\eta$ and the corresponding spectra recorded at $V = \SI{-60}{\volt}$ and $V = \SI{-120}{\volt}$ are shown in Fig.~\ref{fig2}. We observe the direct gap $A$-exciton transition in all spectra at $\sim \SI{1.875}{\electronvolt}$. The degree of circular polarization is derived from the PL $I^{+}$ and $I^{-}$ spectra that is defined by $\eta(E) = (I^{+}(E)-I^{-}(E))/(I^{+}(E) + I^{-}(E))$ and plotted in addition to the PL spectra in Fig.~\ref{fig2}(b). The PL in monolayer MoS$_2$ is strongly circularly polarized with $\eta \sim 80\%$. In general, $\eta$ is a measure for the optically excited steady-state carrier populations in the $K$ and $K^{\prime}$ valleys (schematically shown in Fig.~\ref{fig2}(a)) reflecting the interplay between the \textit{inter}band optical selection rules, as well as \textit{intra}- and \textit{inter}valley relaxation dynamics that occur due to thermalization, and radiative recombination lifetimes.~\cite{Palummo.2015,Robert.2016} For the monolayer, we observe a field-independent $\eta$, which is a direct consequence of the inherently broken inversion symmetry and field-independent \textit{inter}valley scattering.~\cite{Wu.2013} As evident from the data, $\eta$ for monolayer MoS$_2$ is constant throughout the whole range of applied gate-voltages with a high value of $|\eta^{max}_{1L}| = 0.8$ in good agreement with previous reports.~\cite{Heinz2012,Cui2012,Wu.2013,Kiseoglu}

%

\begin{figure*}
	\scalebox{\figurescale}{\includegraphics[width=1\linewidth]{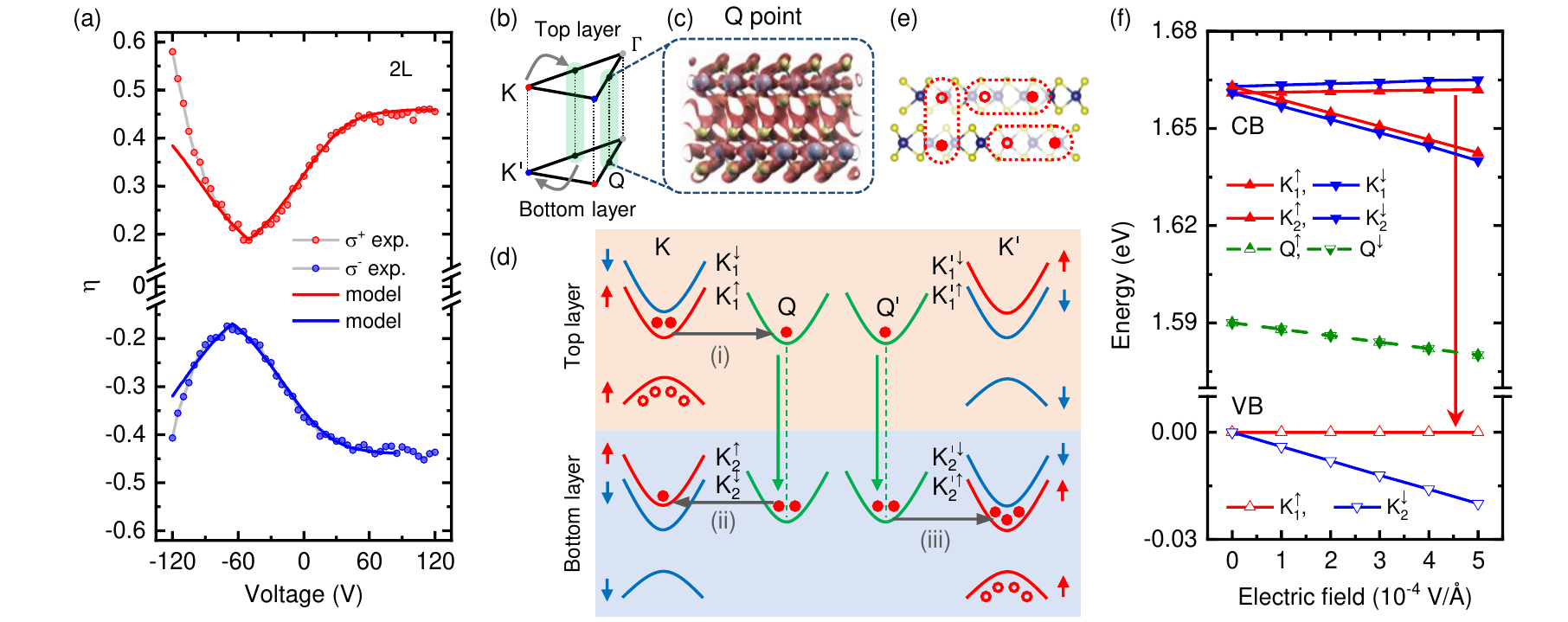}}
	\renewcommand{\figurename}{FIG.|}
	\caption{\label{fig3}
		(a) Tuning of orbital \textit{inter}layer coupling in bilayer MoS$_2$. Voltage controlled degree of valley polarization for $\sigma^+$ (red circles) and $\sigma^-$ (blue circles) excitation. The solid lines is the calculated $\eta$ from a rate-equation model.
		(b) First Brillouin zone and high symmetry points of the bilayer MoS$_2$.
		(c) $Ab$-$initio$ calculated wave function distribution at the $Q$ point is strongly delocalized between top and bottom layer.
		(d) Schematic of the rate equation model used for describing the voltage controlled $\eta$. Valleys in red (blue) correspond to spin-up (spin-down) states. $Q$ and $Q^{\prime}$ valleys are distributed through both layers, as indicated by the dotted line. Grey arrows represent the electronic transitions promoted by the gate-voltage and generates an electric field pointing from bottom to top layer. The green vertical arrows represent the polarization of the $Q$ valleys that pushes the electrons from top layer to bottom layer. The CB electronic population is represented by red circles and holes are represented with open circles.
		(e) Schematic of intra- and \textit{inter}layer excitons in the bilayer MoS$_2$.
		(f) $ab$-$initio$ calculated changes of the high symmetry points $K$/$K^{\prime}$ and $Q$/$Q^{\prime}$ as a function of an external applied electric field $F$ for the top and bottom layer.
		}
\end{figure*}

Unlike the monolayer, the inversion symmetric bilayer shows a strongly tunable $\eta$ (see Fig.~\ref{fig3}(a)). Here, we observe the lowest $\eta = 0.2$ for $V = \SI{-60}{\volt}$ where crystal symmetry is maximally restored in our asymmetric device in very good agreement with the absence of field-activated Raman modes (see Fig.~\ref{fig1}(b)). In contrast, for $V = \SI{-120}{\volt}$, symmetry is maximally broken in our device for which we also observe the field-activated Raman modes with highest intensity. This observation clearly reflects the lifted crystal inversion symmetry, and therefore change in the electronic \textit{inter}layer coupling directly affecting the degree of valley polarization in the bilayer.

Considering the weak \textit{inter}layer coupling in TMDCs, relaxation dynamics are expected to significantly depend on the hybridization of sulfur $p$-orbitals between proximal layers in MoS$_2$. While the conduction bands at $K$/$K^{\prime}$ show only a weak admixture of the sulfur $p$-orbitals since they are mainly comprised of Mo $d$-orbitals, the $Q$ points have a predominant contribution from the $p$-orbitals.~\cite{Liu.2013} The relevant high symmetry points in momentum space of the coupled bilayer system are schematically depicted in Fig.~\ref{fig3}(b). The $Q$ points strongly connect the top and bottom layer, which is also directly apparent from the strong delocalization of our DFT-computed wave function at the $Q$ point (see Fig.~\ref{fig3}(c)). Note that holes in multilayer MoS$_2$ also show a delocalization between $K$ points,~\cite{Gong2013,Deilmann.2018,Slobodeniuk.2019,Guo.2018,Hsu.2019, Leisgang.2020} however, this delocalization is much weaker as compared to the strong delocalization of electrons in the CB at $Q$ (see Supplemental Material for additional wave function calculations at the $K$ point).

Our experiments show that even when the excitation is circularly polarized and resonant with the $A$-exciton, a fraction (percentage $s$) of light is emitted with opposite helicity (at $\sim\SI{-60}{\volt}$). This recombination channel implies an \textit{inter}valley scattering of carriers that is independent of the applied gate-voltage and should be also present in multilayered structures. The variation of $\eta$ as function of the gate-voltage is an effect that can only be observed in multilayered systems and is the result of an interplay between the realignment of the hybridized bands and  transfer of photoexcited carriers through different layers.

To explain the degree of polarization in the PL emission of multilayered systems, we consider only the $\sigma^+$ excitation and a voltage (electric field) that pushes the electrons from the top layer to the bottom layer (see Fig.~\ref{fig3}(d)). Since this model is symmetric, electric fields in the opposite directions and $\sigma^-$ excitation are equivalent. The quasi-resonant excitation of a bilayer MoS$_2$ generates a steady-state electronic distribution in the CB with electrons residing in the $Q$, $Q^{\prime}$, $K$ and $K^{\prime}$ valleys (Fig.~\ref{fig3}(d)).~\cite{Steinhoff.2014,Steinhoff.2015,Steinhoff.2016,Wallauer.2016,Mado.2020} As we consider only the scattering of electrons from $Q$ to $K$ valley and from $Q^{\prime}$ to $K^{\prime}$ valley, we analyze each case separately. Considering that we excited $n_0$ electrons in each case, a fraction $c$ of them reside in the $K_{1}^{\uparrow}$ valley of the top layer, while $(1-c) n_0$ are in the $Q$ valley, distributed through both layers (see dotted lines Fig.~\ref{fig3}(d)). Without a gate-voltage applied, the differential rate equations that describe the dynamics of these two levels are
\begin{equation}
\label{eq1}
    \dot{n}_{K_{1}^{\uparrow}} = c n_0 - \frac{n_{K_{1}^{\uparrow}}}{\tau_{K}} - \frac{n_{K_{1}^{\uparrow}}}{T_{K}}
\end{equation}
and 
\begin{equation}
\label{eq2}
    \dot{n}_{Q} = (1-c) n_0 - \frac{n_{Q}}{T_{Q}},
\end{equation}
where $n_{X}$ is the electronic population in the $X$ state and $\dot{n}_{X}$ its time derivative. $\tau_{K}$ is the radiative recombination lifetime and, $T_{K}$ and $T_{Q}$ are the lifetime for non-radiative recombination channels in each valley, $K$ and $Q$, respectively.

\begin{figure*}
	\scalebox{\figurescale}{\includegraphics[width=1\linewidth]{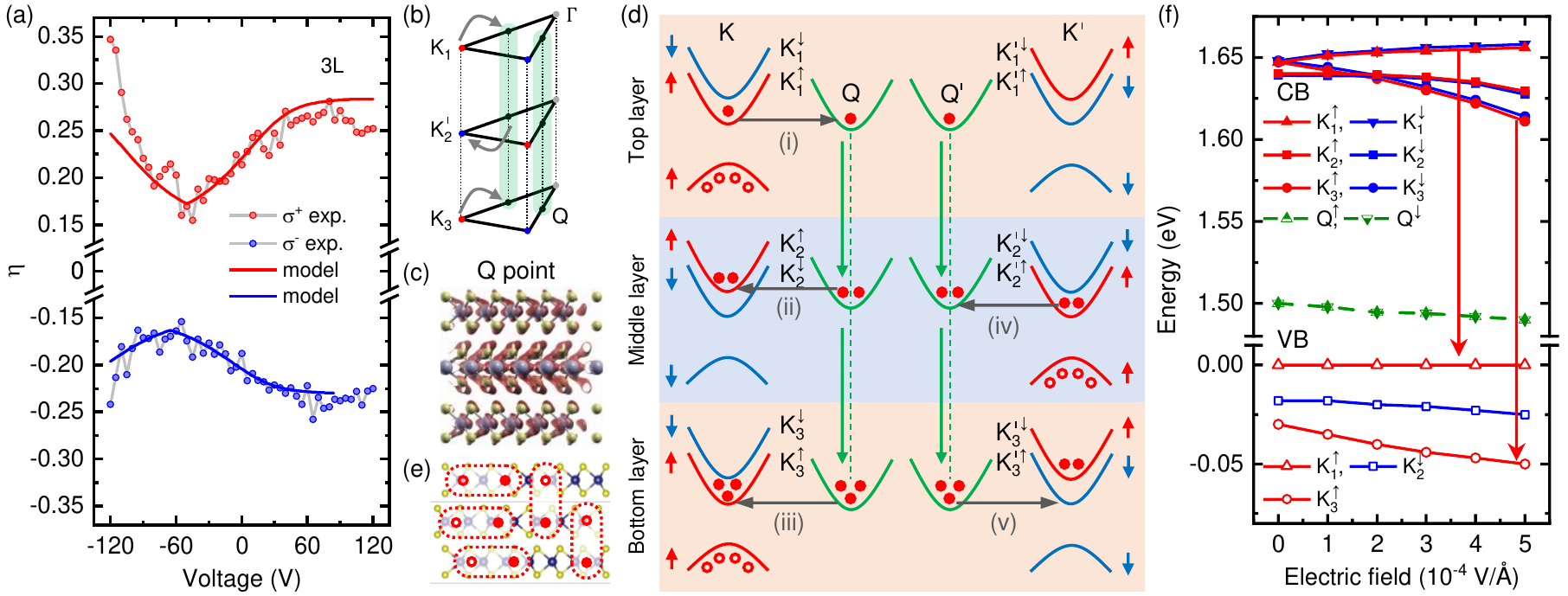}}
	\renewcommand{\figurename}{FIG.|}
	\caption{\label{fig4}
		(a) Tuning of orbital \textit{inter}layer coupling in trilayer MoS$_2$. Voltage controlled degree of valley polarization for $\sigma^+$ (red circles) and $\sigma^-$ (blue circles) excitation. The solid lines are the calculated rate-equation model.
		(b) First Brillouin zone and high symmetry points of the trilayer MoS$_2$.
		(c) $ab$-$initio$ calculated wave function distribution at the $Q$ point is strongly delocalized between top, middle and bottom layer.
		(d) Schematic of the rate equation model used for describing the voltage controlled $\eta$.
		(e) The trilayer hosts \textit{intra}- and \textit{inter}layer excitons.
		(f) $ab$-$initio$ calculated changes of the high symmetry points $K$/$K^{\prime}$ and $Q$ as a function of an external applied electric field $F$ for the top, middle and bottom layer.
		}
\end{figure*}

The application of a gate-voltage generates additional transitions that modify the steady-state and consequently the polarization of the photoluminescence. These transitions are labeled as (i) and (ii) in Fig.~\ref{fig3}(d) for the $Q$ to $K$ scattering case. The green arrows represent the polarization of the $Q$ valley that facilitates the transfer of electrons to the bottom layer. This is the central effect in this model affecting the circular polarization degree for all multilayer samples. The redistribution of electrons via the $Q$ point manifests in the formation of \textit{intra}- and \textit{inter}layer excitons in the bilayer (see Fig~\ref{fig3}(e)). The field-dependent first principle calculations and corresponding energy shifts of the $K$ and $Q$ points in Fig.~\ref{fig3}(f) show, as the most important effect, a strong reduction of the energy distance between the $K^{\uparrow}_2$ point and the $Q$ valley. This enhances transition (ii) at the same time that transition (i) appears to have, in comparison, a reduced contribution. On the other hand, the outstanding effect of the gate-voltage is the population increase of electrons in bright states ($K$ point) at the expense of dark states ($Q$ valley) and this effect can be well described by subtracting $A(V) n_{Q}/T_{QK}$ from the $Q$ valley rate equation. Here, $A(V)$ accounts for the tuneability of the system with gate-voltage $V$, and $T_{QT}$ represents the lifetime for transition (ii). This term constitutes in a source term for the latter level, whose complete rate equation is
\begin{equation}
\label{eq3}
    \dot{n}_{K_{2}^{\uparrow}} = A(V) \frac{n_{Q}}{T_{QK}} - \frac{n_{K_{2}^{\uparrow}}}{\tau_{K}^{\prime}} - \frac{n_{K_{2}^{\uparrow}}}{T_{K}^{\prime}}.
\end{equation}

We model the function $A(V)$ as an exponential growth
\begin{equation}
\label{eq4}
    A = e^{\frac{V-V_0}{V_0}}-e^{-1},
\end{equation}
\noindent
whose value is identically zero for $V = \SI{0}{\volt}$ and $V_0$ is a fitting parameter. It is important to note that the interlayer exciton in MoS$_2$ is the result of a strong admixture between a $B$ intralayer exciton and an $A$ interlayer exciton.~\cite{Leisgang.2020, Gerber.2019, Deilmann.2018} In our case, we do not consider this situation because the laser energy used in the experiments does not allow for the formation of $B$ excitons. Holes are strongly localized in the top layer and the recombination process from the $K_{2}^{\uparrow}$ level would describe the annihilation of \textit{inter}layer excitons with radiative and non-radiative lifetimes of $\tau_{K^{\prime}}$ and $T_{K^{\prime}}$, respectively. 

In the scattering from $Q^{\prime}$ and $K^{\prime}$ case, the excitation with $\sigma^+$ light generates spin-up electrons in $Q^{\prime}$ of the bilayer and $K_{2}^{\prime {\uparrow}}$ of the bottom layer with similar dynamics as in the $Q$ and $K$ points. Including the gate-voltage effect which activates the transition labeled (iii) in Fig.~\ref{fig3}(d), their rate equations are
\begin{equation}
\label{eq5}
    \dot{n}_{K_{2}^{\prime {\uparrow}}} = c n_0 - \frac{n_{K_{2}^{\prime {\uparrow}}}}{\tau_{K}} - \frac{n_{K_{2}^{\prime {\uparrow}}}}{T_{K}} + A(V) \frac{n_{Q^{\prime}}}{T_{QK}}
\end{equation}
and 
\begin{equation}
\label{eq6}
    \dot{n}_{Q^{\prime}} = (1-c) n_0 - \frac{n_{Q^{\prime}}}{T_{Q}} - A(V) \frac{n_{Q^{\prime}}}{T_{QK}}.
\end{equation}

Since both layers are equivalent, the radiative and non-radiative lifetimes involved in the process are the same. In addition, $Q$ and $Q^{\prime}$ are also equivalent points of the bilayer with the same non-radiative lifetime. 

All the radiative transitions described in the rate equations above correspond to light emitted with $\sigma^+$ polarization. The \textit{inter}valley scattering, $s$, yields $\sigma^-$ polarized light. By solving these equations in the steady-state with realistic material parameters, we determine the $\sigma^+$ polarized emission intensity and calculate the circular polarization degree of the PL using $c$, $s$ and $V_0$ as fitting parameters. Note that this emission originates from both, \textit{intra}- and \textit{inter}layer excitons (see Fig.~\ref{fig3}(e)). The total $\sigma^+$ emission is then the sum of both processes. The calculated electric field dependent circular polarization degree is presented in Fig.~\ref{fig3}(a), showing excellent agreement with the experiment. Both data sets are fitted individually due to a small difference in the voltage offset that originates from the subsequent recording of data and the hysteresis of the device.~\cite{Klein.2016} The corresponding material and fitting parameters are listed in the Supplemental Material.

Since our model stems from steady-state electron distributions, similar dependencies of $\eta$ are expected for systems with more than two layers. Indeed, probing the trilayer MoS$_2$, we obtain a similar modulation of $\eta$ as shown in Fig.~\ref{fig4}(a). For the case where the symmetry between the layers is restored ($\SI{-60}{\volt}$), we obtain a minimum $\eta \sim 0.15$ while for the highest electric-field ($\SI{-120}{\volt}$) a maximum $\eta \sim 0.35$ is observed. Similar to the bilayer case, we find a strong delocalization of the electron wave function at the $Q$ points (see Fig.~\ref{fig4}(b) and (c)) suggesting a similar steady-state population redistribution of electrons throughout the trilayer. The level scheme for a trilayer is analogous to the bilayer by adding an additional layer as shown in Fig.~\ref{fig4}(d). Similar to the bilayer, it allows the formation of \textit{intra}- and \textit{inter}layer excitons (see Fig.~\ref{fig4}(e)). The computed CB and VB edges of $K$/$K^{\prime}$ and $Q$/$Q^{\prime}$ points are shown in Fig.~\ref{fig4}(f). For a finite field of $> 2 \cdot 10^{-4}\SI{}{\volt\per\angstrom}$, the energy sequence in the CB and VB follow the stacking order of the layers resulting in a field-dependent electron distribution throughout the trilayer that is mediated via the $Q$ point (see Fig.~\ref{fig4}(d) and (f)). We solved the rate equation model for a trilayer system (see Fig.~\ref{fig4}(d)) analogously to the bilayer. The resulting $\eta$ is again in excellent agreement with experimental data (see Fig.~\ref{fig4}(a)). 

The valley optical selection rules in TMDCs are strongly connected to the crystal inversion symmetry and time-reversal symmetry (Kramer's degeneracy). 2H trilayer TMDCs are intrinsically inversion symmetry broken and therefore, tuneability of the optical valley dichroism is not expected. This results for example in a non-tunable second-harmonic generation.~\cite{Klein.2016b} However, our observation of field-tunable $\eta$ in the trilayer suggests that the breaking of the Kramer's degeneracy and the accompanying occurrence of a valley magnetic moment~\cite{Wu.2013} is not sufficient to explain our data (see Fig.~\ref{fig4}(a)). Our results and microscopic modeling based on DFT calculation input suggest that the high-symmetry $Q$ point and a field-dependent change in the \textit{inter}layer hybridization, and therefore \textit{inter}layer hopping can strongly modify the steady-state electron, and therefore exciton population in the bi- and trilayer system manifesting in the tunable valley dichroism. This is in contrast to earlier work that attributed it solely to a change in the valley magnetic moment.~\cite{Wu.2013,Kormnyos.2018}

\section{Conclusion}
In summary, we demonstrated the electrical control of \textit{inter}layer vibrational and orbital coupling in bi- and trilayer MoS$_2$. Field-dependent Raman spectroscopy is a useful tool to probe the symmetry and layer decoupling of bi- and trilayer crystals. Our work suggests that the $Q$ point in multilayer TMDCs is important for understanding steady-state electron populations and, therefore, the spin-valley dichroism of TMDCs and other vdW heterostructures.

%
%
\section{Acknowledgements}
We gratefully acknowledge financial support from ExQM PhD programme of the Elite Network of Bavaria, the German Excellence Initiative via the Nanosystems Initiative Munich (NIM), the Deutsche Forschungsgemeinschaft (DFG) via the clusters of excellence e-conversion (EXC 2111) and MCQST (EXC 2089), via SPP 2244, the European Union’s Horizon 2020 research and innovation programme under grant agreement No. 820423 (S2QUIP) and through the TUM International Graduate School of Science and Engineering (IGSSE). J.K. acknowledges support by the Alexander von Humboldt foundation.
A.K. acknowledges DFG project GRK 2247/1 (QM3), and the association with SPP 2244. A.K. and T.B. acknowledge the support by the DFG within SFB 1415 and the high-performance computing center of ZIH Dresden for computational resources. We further thank Ursula Wurstbauer and Andor Korm\'{a}nyos for insightful and stimulating discussions.\newline

\section{Author contributions}

J.K., J.W., K.M. and J.J.F. conceived and designed the experiments, J.K. and J.W. prepared the samples, J.K. and J.W. performed the optical measurements, J.K. analyzed the data, P.S., J.K. and A.V.S. developed the rate equation model, A.K. and L.M. computed the electric field-dependent Raman and phonon-dispersion spectra, T.B. computed the field-dependent electronic band structures, J.K. wrote the paper with input from all co-authors. All authors reviewed the manuscript.

\section{Methods}

\subsection{Device fabrication}
The MoS$_2$ crystal is mechanically exfoliated onto a $\SI{295}{\nano\meter}$ thick SiO$_2$ layer on a heavily n-doped silicon substrate. Spatially resolved Raman spectroscopy and white light interferometry are used to identify mono-, bi- and trilayer regions of the flake, before it is capped with a \SI{20}{\nano\meter} thick Al$_2$O$_3$ dielectric using atomic layer deposition (ALD) at a low temperature of $\SI{120}{\celsius}$. An electrical contact is established to the silicon substrate and the sample is completed with a \SI{5}{\nano\meter} thick semi-transparent titanium top contact that facilitates optical access to the crystal while tuning the gate-voltage (electric field) in the range $\pm\SI{120}{\volt}$. The device fabrication procedure is found to have no significant deleterious impact on the photoluminescence properties of the atomically thin flake. Further details of the fabrication, electrical characterization and control of the $A$-exciton emission energies using the DC-Stark effect and control of second-harmonic generation (SHG) can be found in Ref.~\cite{Klein.2016} and Ref.~\cite{Klein.2016b}. 

\subsection{Optical spectroscopy}

For low-temperature confocal $\mu$-PL and Raman measurements we keep the device under vacuum in a helium flow cryostat with a lattice temperature kept at $T = \SI{10}{\kelvin}$. For circularly polarized measurements, we use a HeNe laser with an excitation energy of $\SI{1.96}{\electronvolt}$. An excitation power density of $\sim\SI{10}{\kilo\watt\per\centi\meter\squared}$ is used. The laser spot has a diameter of $d_{spot} \sim \SI{1.2}{\micro\meter}$.

\subsection{DFT calculations}

All materials (1L-3L MoS$_2$) were fully optimized, including the lattice parameters and the atomic positions using density-functional theory with Perdew-Burke-Ernzerhof (PBE)\cite{PBE} exchange-correlation functional together with D3 London dispersion correction\cite{D3} as implemented in Crystal17.\cite{Crystal17, Crystal17a} The resulting lattice parameters $a$ and $b$ are: 3.137~\AA\ for 1L, 3.136~\AA\ for 2L, and 3.135~\AA\ for 3L, while interlayer distance between metal centers is $d=5.956$~\AA.
For Mo atoms, we used small-core effective-core pseudopotential (ECP) of the Hay–Wadt type\cite{HW} which accounted for the electrons 1s$^2$–3d$^{10}$, while for S atoms, we used 86-311G$^*$.\cite{S-basis}
The 8$\times$8$\times$1 $k$-point grid was used for structural relaxation. For Raman and phonon-dispersion calculations, geometries were re-optimized applying external electric field of a given strength perpendicular to the basal plane of the layers.
Phonon dispersion calculations were performed using finite displacement method on the 5$\times$5$\times$1 supercells. Raman active modes were calculated at the $\Gamma$-point.

Using the geometries as relaxed with Crystal17, the field-dependent electronic band structures and the corresponding spin projections have been calculated within the framework of DFT as implemented in the {\sc Quantum ESPRESSO} package.\cite{QE1,QE2} We employed full-relativistic, projector-augmented wave potentials\cite{kjpaw} of the version 1.0.0 of the pslibrary\cite{dalcorso2014,*pslib} and we also chose the PBE\cite{PBE} for the exchange-correlation energy. A cutoff of $60\,\mathrm{Ry}$ and $480\,\mathrm{Ry}$ ($1\,\mathrm{Ry}\approx13.6\,\mathrm{eV}$) for the wave functions and the charge density, respectively, was used. The convergence was checked with higher cutoffs which lead to only slight shifts of the semi-core states at energies lower than $\approx-30\,\mathrm{eV}$ below the Fermi energy. The Brillouin zone integration was performed with a $\Gamma$-centered Monkhorst-Pack grid\cite{monkhorst1976} of $18\times18\times1$ $k$-points together with a Gaussian broadening of $0.005\,\mathrm{Ry}$. The self-consistent solution of the Kohn-Sham equations was obtained when the total energy changed by less than $10^{-10}\,\mathrm{Ry}$.

\subsection{Rate equation model}

See Supplemental Material.

\section{Abbreviations}
Two-dimensional, 2D; TMDC, transition metal dichalcogenides; \textmu-PL, micro-photoluminescence; DFT, density functional theory; IR, infrared; Conduction band, CB; Valence band, VB;

%
\section{Additional information}
\subsection{Supplementary Information}
Supplemental Material accompanies this paper.

\subsection{Competing financial interests}
The authors declare no competing financial interests.

%
%

\bibliographystyle{apsrev4-1}
\bibliography{full}

\end{document}


\title{Supplemental Material - Electrical control of orbital and vibrational \textit{inter}layer coupling in bi- and trilayer 2H-MoS$_2$}
%
%
\author{J. Klein}
\affiliation{Walter Schottky Institut and Physik Department, Technische Universit\"at M\"unchen, Am Coulombwall 4, 85748 Garching, Germany}
\affiliation{Department of Materials Science and Engineering, Massachusetts Institute of Technology, Cambridge, Massachusetts 02139, USA}
\author{J. Wierzbowski}
\affiliation{Walter Schottky Institut and Physik Department, Technische Universit\"at M\"unchen, Am Coulombwall 4, 85748 Garching, Germany}
\author{P. Soubelet}
\affiliation{Walter Schottky Institut and Physik Department, Technische Universit\"at M\"unchen, Am Coulombwall 4, 85748 Garching, Germany}
\author{A. Kuc}
\affiliation{Abteilung Ressourcen\"okologie, Helmholtz-Zentrum Dresden-Rossendorf, Forschungsstelle Leipzig, Leipzig, Germany}
\affiliation{Department of Physics \& Earth Science, Jacobs University Bremen, Bremen, Germany}
\author{T. Brumme}
\affiliation{Wilhelm-Ostwald-Institute for Physical and Theoretical Chemistry, Leipzig University, Leipzig, Germany}
\affiliation{Theoretical Chemistry, TU Dresden, Dresden, 01062 Germany}
\author{L. Maschio}
\affiliation{Dipartimento di Chimica and Centre of Excellence NIS (Nanostructured Interfaces and Surfaces), Universit\`a di Torino, via P. Giuria 5, I-10125 Turin, Italy}
\author{K. M\"uller}
\affiliation{Walter Schottky Institut and Physik Department, Technische Universit\"at M\"unchen, Am Coulombwall 4, 85748 Garching, Germany}
\author{A. V. Stier}
\affiliation{Walter Schottky Institut and Physik Department, Technische Universit\"at M\"unchen, Am Coulombwall 4, 85748 Garching, Germany}
\author{J. J. Finley}
\affiliation{Walter Schottky Institut and Physik Department, Technische Universit\"at M\"unchen, Am Coulombwall 4, 85748 Garching, Germany}
%
%
\date{\today}
%

%
\maketitle
%
%

\tableofcontents

\newpage

\section{Experimental setup to probe the circular polarization}

%
\begin{figure}[ht]
\scalebox{\figurescale}{\includegraphics[width=0.7\linewidth]{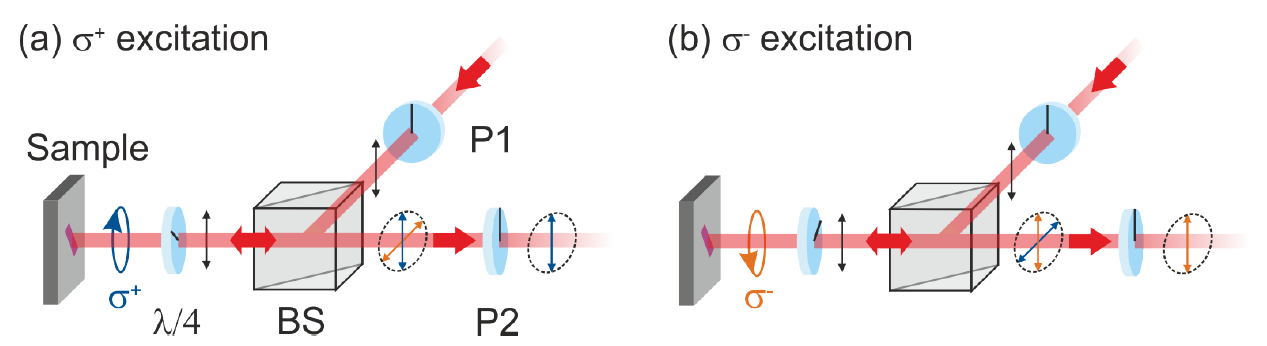}}
\renewcommand{\figurename}{SI Figure}
\caption{\label{SIfig1}
Experimental setup for polarization resolved PL measurements. (a) Experimental configuration for $\sigma^{+}$ polarized excitation. The incident laser beam is linearly polarized with an excitation polarizer (P1). The light passes the beamsplitter (BS) and a quarter wave plate converts $\sigma^{+}$ light. Helical light from the sample is converted into linear light again by passing the quarter waveplate a second time. The detection polarizer (P2) either cuts on parallel or cross-polarized light to analyze the amount of $\sigma^{-}$ and $\sigma^{+}$ polarized light. (b) Same setup but for excitation with $\sigma^{-}$ polarized light.
}
\end{figure}
%

\newpage

\section{Gate-voltage dependent Raman spectra of bi- and trilayer MoS$_2$}

%
\begin{figure}[ht]
\scalebox{\figurescale}{\includegraphics[width=1\linewidth]{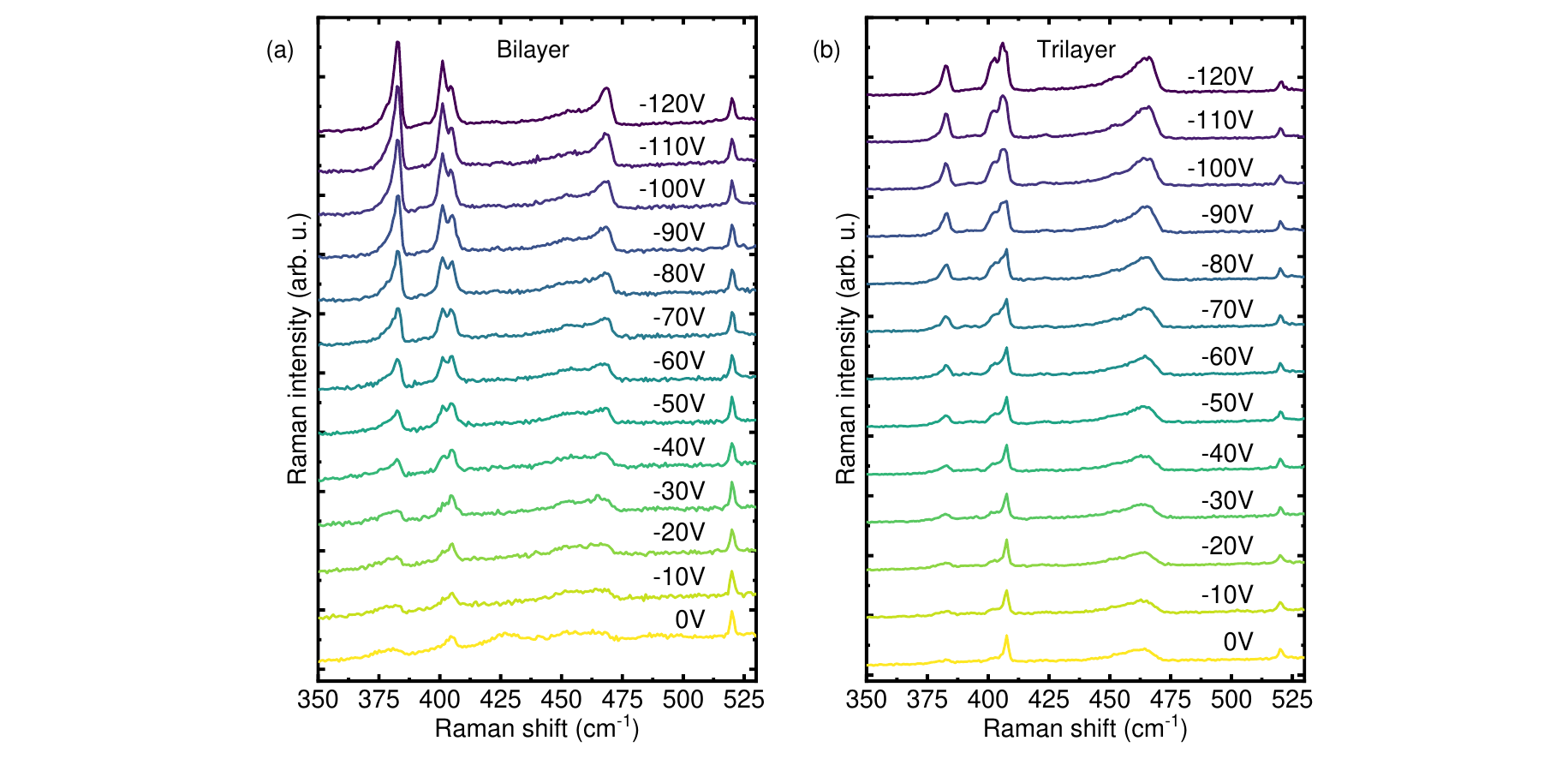}}
\renewcommand{\figurename}{SI Figure}
\caption{\label{SIfigrawraman}
Raman spectra of (b) bilayer and (c) trilayer MoS$_2$ as a function of gate-voltage. Data are taken at a lattice temperature of $\SI{10}{\kelvin}$ and with a laser excitation energy of $\SI{1.96}{\electronvolt}$.
}
\end{figure}
%

\newpage

\section{DFT calculated phonon dispersion with external electric field}

%
\begin{figure}[ht]
\scalebox{\figurescale}{\includegraphics[width=0.8\linewidth]{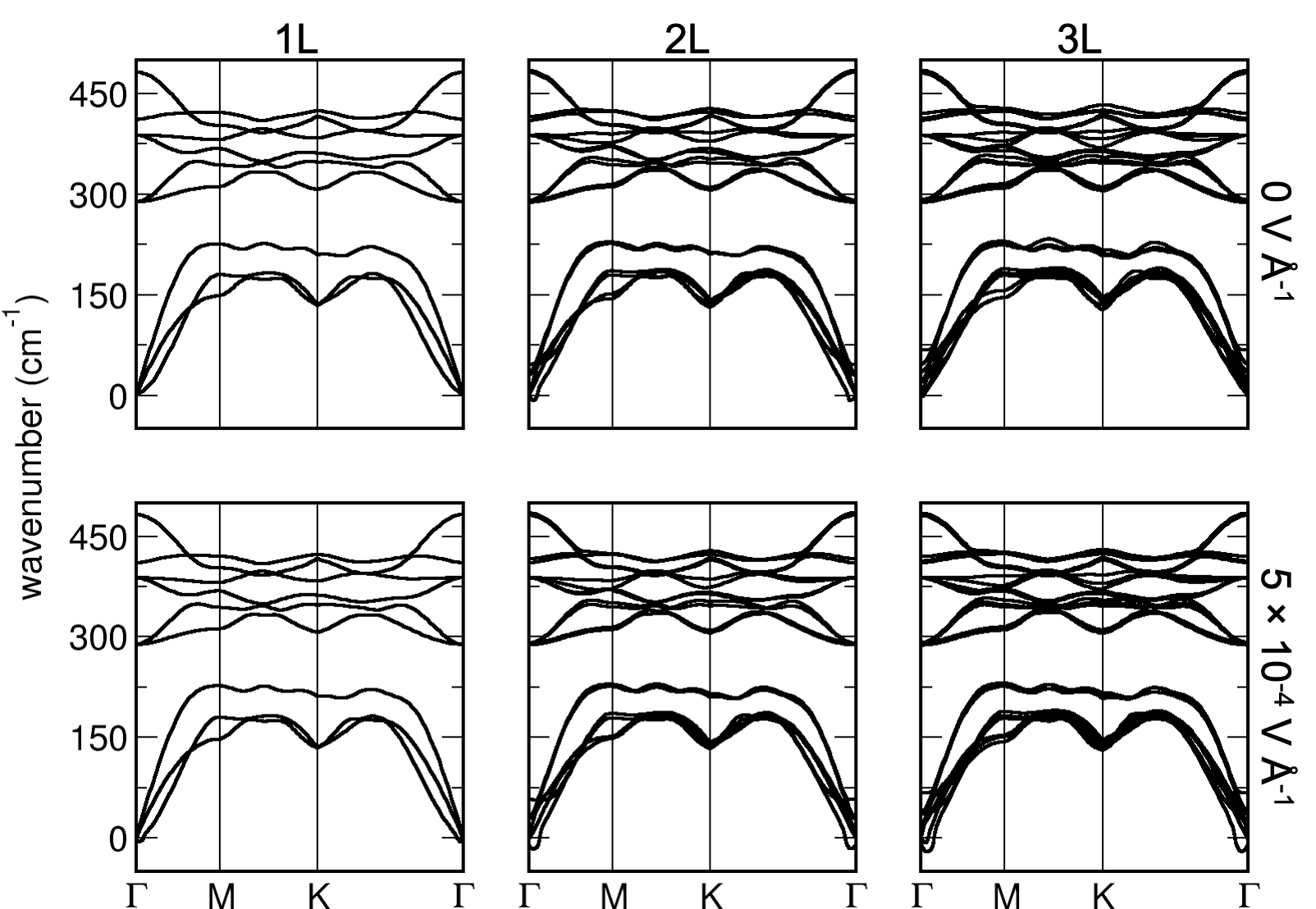}}
\renewcommand{\figurename}{SI Figure}
\caption{\label{SIfig4}
DFT calculated phonon dispersions of mono-, bi- and trilayer MoS$_2$ without and with applied electric field.
}
\end{figure}
%

\newpage

\section{Degree of circular polarization on SiO$_2$/Si}

%
\begin{figure}[ht]
\scalebox{\figurescale}{\includegraphics[width=1\linewidth]{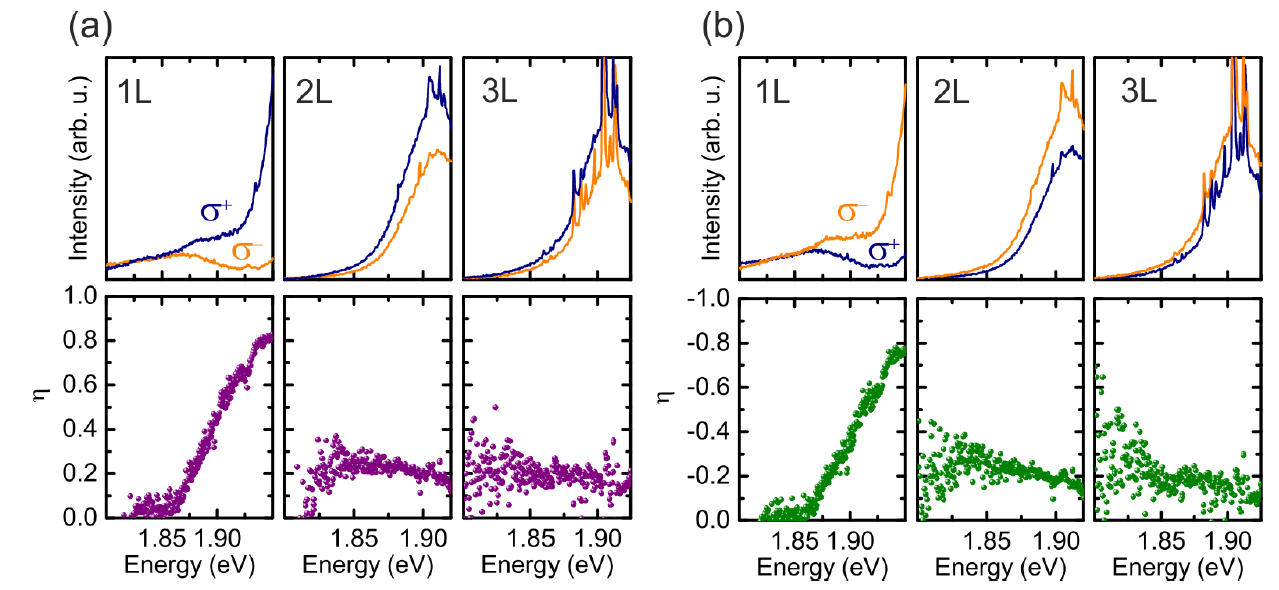}}
\renewcommand{\figurename}{SI Figure}
\caption{\label{SIfig2}
Polarization-resolved spectra obtained by helical light excitation and corresponding
degree of circular polarization of MoS$_2$ on SiO$_2$/Si. (a) Mono-, bi- and trilayer PL spectra obtained by $\sigma^{+}$ excitation and detection of $\sigma^{+}$ (blue) and $\sigma^{-}$ (orange) helical light. The monolayer A exciton is strongly polarization dependent with a degree of circular polarization of $\eta = 0.81$. The low energy defect peak shows no significant polarization dependence. Bi- and trilayer only exhibit a weak polarization dependence with values of $\eta = 0.2$ and $\eta = 0.18$. (b) Polarization-resolved PL spectra for $\sigma^{-}$  excitation. The same quantitative behaviour as for $\sigma^{+}$ excitation is observed. Corresponding values for the degree of circular polarization for mono-, bi- and trilayer are $\eta = -0.78$, $\eta = -0.2$ and $\eta = -0.17$, respectively.
}
\end{figure}
%

\newpage

\section{Degree of circular polarization in the micro-capacitor device}

%
\begin{figure}[ht]
\scalebox{\figurescale}{\includegraphics[width=1\linewidth]{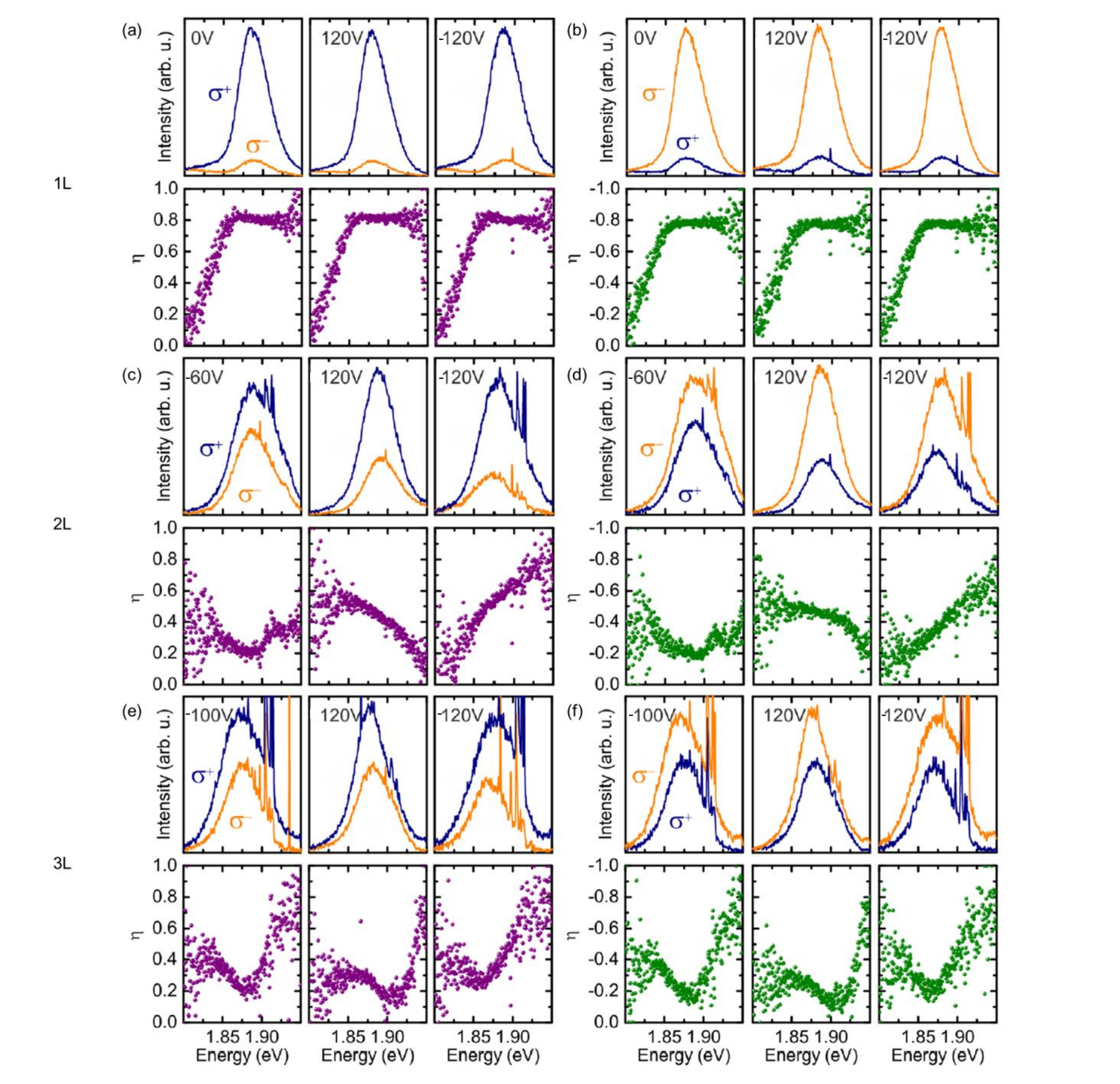}}
\renewcommand{\figurename}{SI Figure}
\caption{\label{SIfig3}
Polarization-resolved spectra of mono-, bi- and trilayer MoS$_2$ in the micro-capacitor device obtained by helical light excitation and corresponding
degree of circular polarization for $\sigma^+$ excitation and $\sigma^-$ detection for (a) monolayer, (c) bilayer and (e) trilayer and $\sigma^-$ and $\sigma^+$ detection for (b) monolayer, (d) bilayer and (f) trilayer MoS$_2$.
}
\end{figure}
%

\newpage

\section{DFT calculated electronic band structure with external electric field}

%
\begin{figure}[ht]
\scalebox{\figurescale}{\includegraphics[width=1\linewidth]{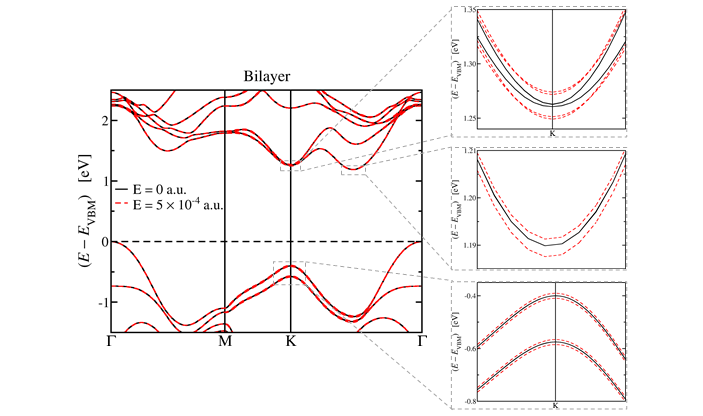}}
\renewcommand{\figurename}{SI Figure}
\caption{\label{SIfig5}
DFT calculated electronic band structure of bilayer MoS$_2$ without (black lines) and with (dashed red lines) an external applied electric field.
}
\end{figure}
%

\newpage

%
\begin{figure}[ht]
\scalebox{\figurescale}{\includegraphics[width=1\linewidth]{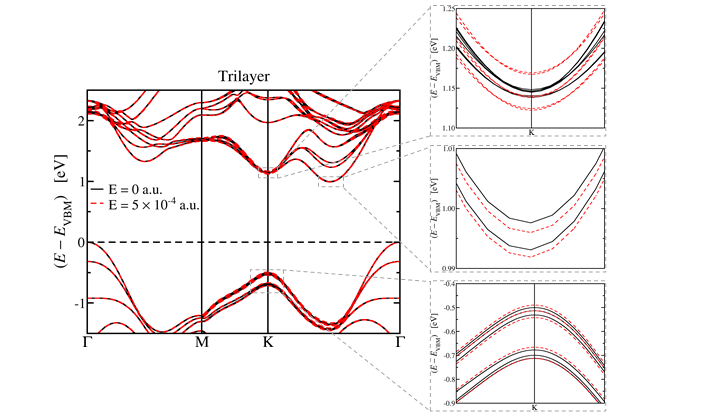}}
\renewcommand{\figurename}{SI Figure}
\caption{\label{SIfig6}
DFT calculated electronic band structure of trilayer MoS$_2$ without (black lines) and with (dashed red lines) an external applied electric field.
}
\end{figure}
%

\newpage

\section{DFT calculated wave functions of valence and conduction bands in bilayer and trilayer MoS$_2$}

%
\begin{figure}[ht]
\scalebox{\figurescale}{\includegraphics[width=1\linewidth]{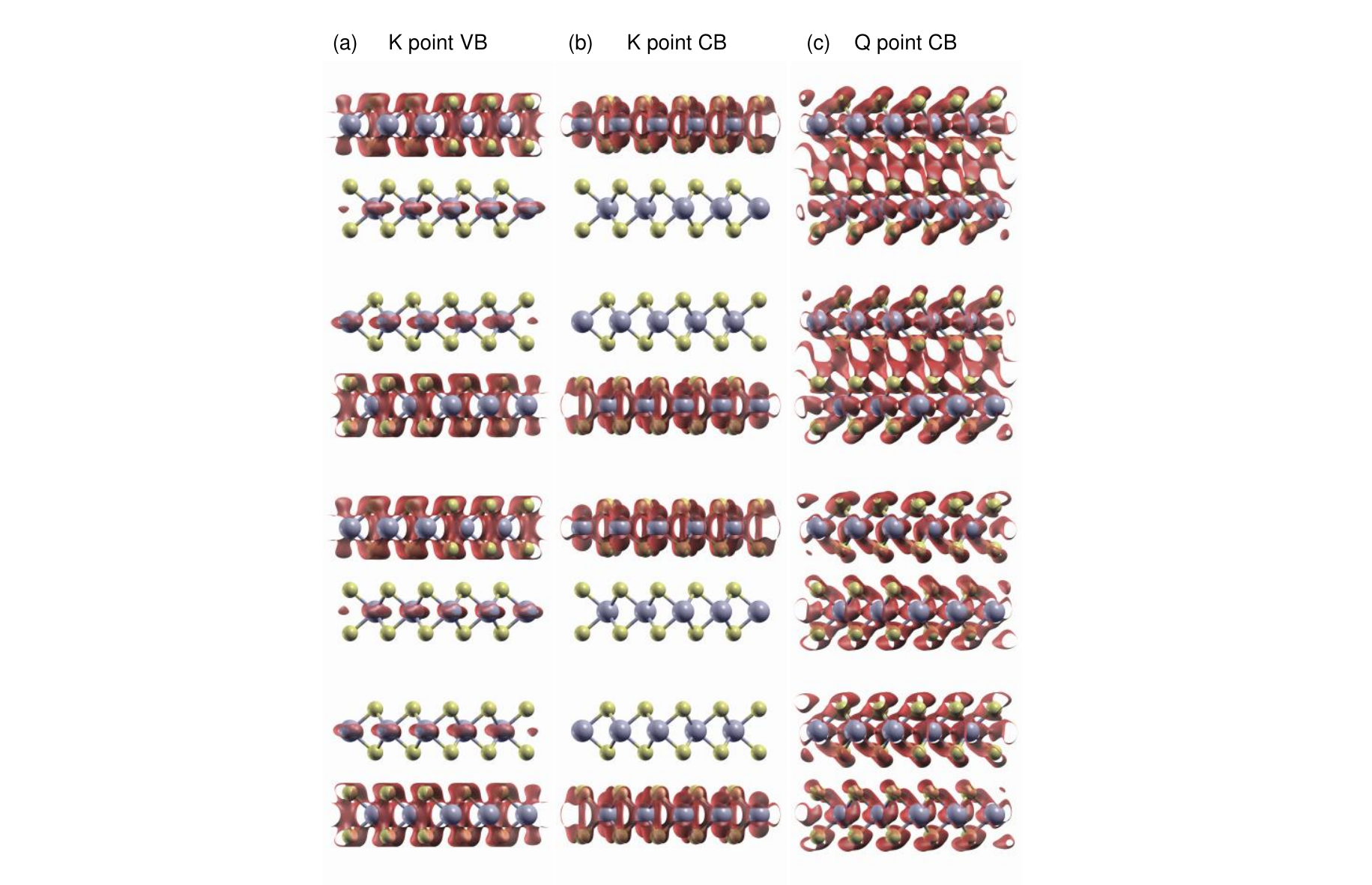}}
\renewcommand{\figurename}{SI Figure}
\caption{\label{SIfigwf2L}
DFT calculated wave function of electrons and holes in bilayer MoS$_2$ without external electric field. (a) Wave function distribution of holes at the $K$ point for the 4 individual valence bands in the bilayer (see SI Fig.~\ref{SIfig5}). (b) Wave function distribution of electrons at the $K$ point for the 4 individual conduction bands. (c) Wave function distribution of electrons at the $Q$ point for the 4 individual conduction bands.
}
\end{figure}
%

%
\begin{figure}[ht]
\scalebox{\figurescale}{\includegraphics[width=0.99\linewidth]{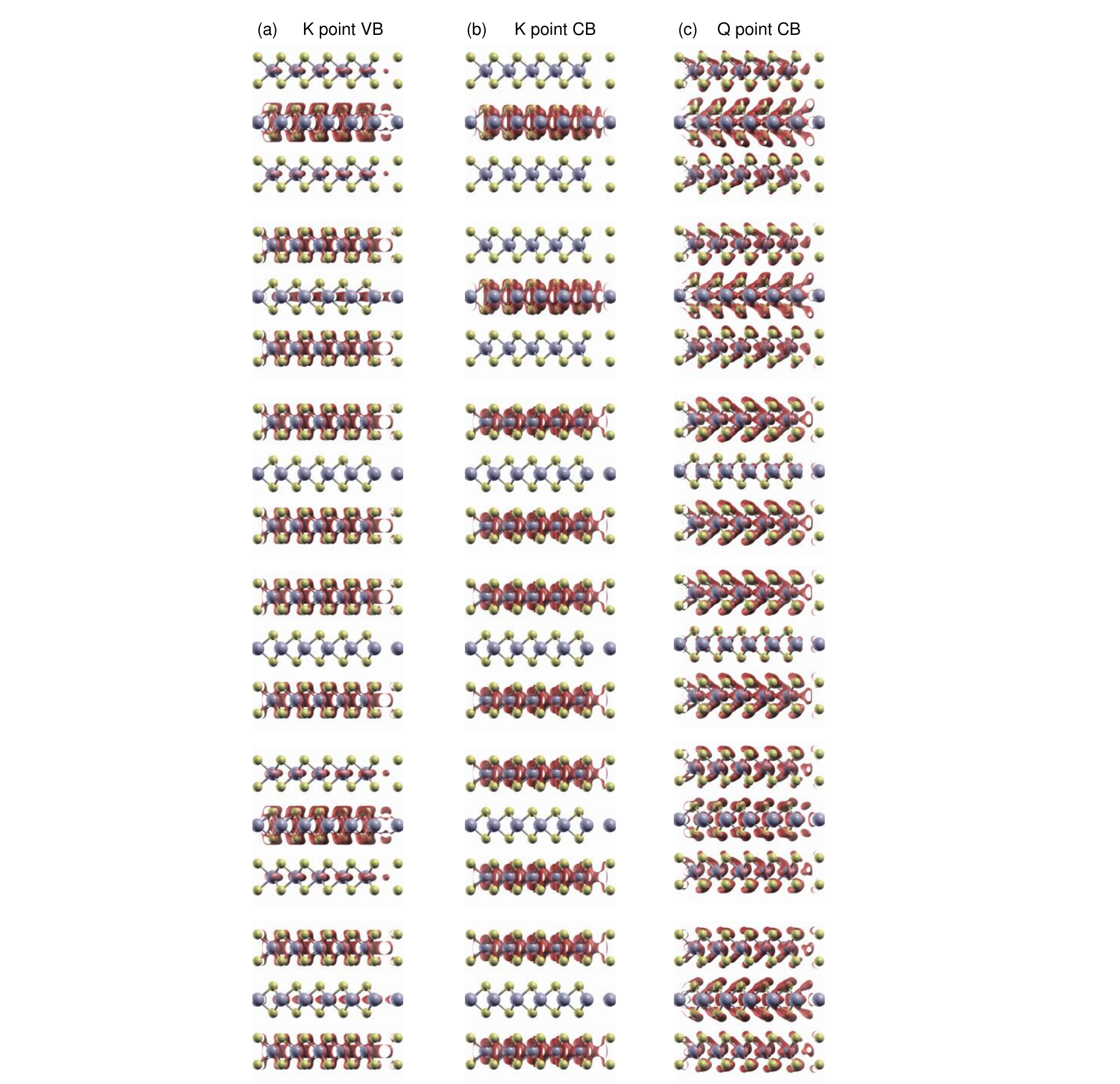}}
\renewcommand{\figurename}{SI Figure}
\caption{\label{SIfigwf3L}
DFT calculated wave function of electrons and holes in trilayer MoS$_2$ without external electric field. (a) Wave function distribution of holes at the $K$ point for the 6 individual valence bands in the trilayer (see SI Fig.~\ref{SIfig6}). (b) Wave function distribution of electrons at the $K$ point for the 6 individual conduction bands. (c) Wave function distribution of electrons at the $Q$ point for the 6 individual conduction bands.
}
\end{figure}
%

\newpage

\section{Rate equation model}

The absorption of photons in a TMDC material promotes electrons from their initial state in the VB into the excited state in the CB. The continuous excitation of the material generates a steady-state in which the population of different valleys in the CB incorporate electrons at the same rate in which the radiative and non-radiative channels send the excited carriers to the VB. The distribution of electrons in the CB bands depends on the energy and polarization of the absorbed photon as much as the relative alignments of the different valleys and their respective relaxation mechanisms~\cite{Mado.2020,Steinhoff.2015}. Affecting the relative alignments between the different valleys in the CB gives rise to the redistribution of the excited electrons, as it is presented in Ref.~\cite{Steinhoff.2015}.

The effect of a gate-voltage in the trilayer system can be described in two different cases that encompass the scattering of electrons from $K$ to $Q$ and from $K^{\prime}$ to $Q^{\prime}$. The rate equations for the first case involve the states $K_{1}^{{\uparrow}}$, $K_{2}^{{\uparrow}}$, $K_{3}^{{\uparrow}}$ and $Q$. Considering that we excite $n_0$ electrons, $1-c$ are located in the $Q$ valley through all the layers and, $K_{1}^{{\uparrow}}$ and $K_{2}^{{\uparrow}}$ acquire $c n_0/2$ each one. Using the same notation we used previously for the electronic population in each state and its derivative in time, their rate equations are
\begin{equation}
\label{eq7}
    \dot{n}_{K_{1}^{\uparrow}} = \frac{c}{2} n_0 - \frac{n_{K_{1}^{\uparrow}}}{\tau_{K}} - \frac{n_{K_{1}^{\uparrow}}}{T_{K}},
\end{equation}
\begin{equation}
\label{eq8}
    \dot{n}_{K_{2}^{\uparrow}} =  A(V) \frac{n_{Q}}{T_{QK}} - \frac{n_{K_{2}^{\uparrow}}}{\tau_{K}^{\prime}} - \frac{n_{K_{2}^{\uparrow}}}{T_{K}^{\prime}},
\end{equation}
\begin{equation}
\label{eq9}
    \dot{n}_{K_{3}^{\uparrow}} = \frac{c}{2} n_0 + A(V) \frac{n_{Q}}{T_{QK}} - \frac{n_{K_{3}^{\uparrow}}}{\tau_{K}} - \frac{n_{K_{3}^{\uparrow}}}{T_{K}}
\end{equation}
and 
\begin{equation}
\label{eq10}
    \dot{n}_{Q} = (1-c) n_0 - \frac{n_{Q}}{T_{Q}} - 2A(V) \frac{n_{Q}}{T_{QK}}.
\end{equation}

For simplicity, we have assumed that the different decay rates are the same we used for the bilayer system and that transitions (ii) and (iii) in Fig.~4 in the main text are mathematically equivalent. As before, we have not included the transition labeled as (i). As in this case holes are strongly localized in the top and bottom layer, the electrons in $K_{2}^{\uparrow}$ may bind holes from the top layer to form \textit{inter}layer excitons between the top and the middle layer. The formation of \textit{inter}layer excitons with holes in the bottom layer are energetically disfavored by the direction of the electric field.

\begin{table}
\begin{ruledtabular}
\begin{tabularx}{\linewidth} {cccc}

 intra-layer radiative lifetime & $\tau_K$ & & 100\,fs \\
 intra-layer non-radiative lifetime & $T_K$ & & 100\,fs \\
 indirect non-radiative lifetime & $T_Q$ & & 2000\,fs \\
 $Q$ to $K$ scattering lifetime & $T_{QK}$ & & 2000\,fs \\
 inter-layer radiative lifetime & $\tau'_K$ & & 300\,fs \\
 inter-layer non-radiative lifetime & $T'_K$ & & 300\,fs \\
\hline
\hline
  & & $\sigma^+$  & $\sigma^-$ \\
\hline
\multirow{3}{*}{2L}       & $c$ & 0.37 & 0.37 \\
                    & $s$ & 40\% & 41\% \\
                    & $V_0$ & 33\,V & 33\,V \\
\hline 
\multirow{3}{*}{3L}       & $c$ & 0.77 & 0.78 \\
                    & $s$ & 45\% & 45\% \\
                    & $V_0$ & 33\,V & 33\,V \\
\end{tabularx}
\caption{Material and fitting parameters used in the rate equation model.}
\label{table:values_param}
\end{ruledtabular}
\end{table}

Finally, we analyze the scattering of electrons from the $Q^{\prime}$ to the $K^{\prime}$ valley in the trilayer system. In this case, $\sigma^+$ polarized light populates only the $K^{\prime \uparrow}_2$ and the $Q^{\prime}$ valleys with $c n_0/2$ and $1-c n_0$ electrons, respectively, and, as consequence, the total number of excited electrons is smaller than $n_0$. The rate equations in this case are
\begin{equation}
\label{eq11}
    \dot{n}_{K_{2}^{\prime\uparrow}} = \frac{c}{2} n_0 - \frac{n_{K_{2}^{\prime\uparrow}}}{\tau_{K}} - \frac{n_{K_{2}^{\prime\uparrow}}}{T_{K}},
\end{equation}
\begin{equation}
\label{eq12}
    \dot{n}_{Q^{\prime}} = (1-c) n_0 - \frac{n_{Q}}{T_{Q}} - A(V) \frac{n_{Q^{\prime}}}{T_{QK}}
\end{equation}
and
\begin{equation}
\label{eq13}
    \dot{n}_{K_{3}^{\prime\uparrow}} = A(V) \frac{n_{Q^{\prime}}}{T_{QK}} - \frac{n_{K_{3}^{\prime\uparrow}}}{\tau^{\prime}_{K}} - \frac{n_{K_{3}^{\prime\uparrow}}}{T^{\prime}_{K}}.
\end{equation}

As before, we did not take into account the transition labeled as (iv) in Fig.~4 in the main text. In this latter case, the electrons in $K_{3}^{\prime\uparrow}$ level would bind holes located in the middle layer, forming \textit{inter}layer excitons between the middle and the bottom layer.  

The degree of circular polarization in the photoluminescence is obtained in the same way as in the bilayer case and is presented by the solid curve in Fig.~4(a) in the main text. Material and fitting parameters are presented in table~\ref{table:values_param}.\newline

%
%
%
%
\bibliographystyle{apsrev}
\bibliography{full}